\documentclass{article}
\usepackage{graphicx}
\usepackage{amsmath}
\usepackage{amsfonts}
\usepackage{amssymb}
\newtheorem{theorem}{Theorem}

\begin{document}

\title{Two-dimensional models\\{\small Contribution proposed for the project ``Encyclopedia of Mathematical Physis''}}
\author{Bert Schroer\\CBPF, Rua Dr. Xavier Sigaud 150 \\22290-180 Rio de Janeiro, Brazil\\and Institut fuer Theoretische Physik der FU Berlin, Germany}
\date{April 2005}
\maketitle

\section{History and motivation}

Local quantum physics of systems with infinitely many interacting degrees of
freedom leads to situations whose understanding often requires new physical
intuition and mathematical concepts beyond that acquired in quantum mechanics
and perturbative constructions in quantum field theory. In this situation
two-dimensional soluble models turned out to play an important role. On the
one hand they illustrate new concepts and sometimes remove misconceptions in
an area where new physical intuition is still in process of being formed. On
the other hand rigorously soluble models confirm that the underlying physical
postulates are mathematically consistent, a task which for interacting systems
with infinite degrees of freedom is mostly beyond the capability of pedestrian
methods or brute force application of hard analysis on models whose natural
invariances has been mutilated by a cut-off.

In order to underline these points and motivate the interest in 2-dimensional
QFT, let us briefly look at the history, in particular at the physical
significance of the three oldest two-dimensional models of relevance for
statistical mechanics and relativistic particle physics, in chronological
order: the Lenz-Ising model, Jordan's model of bosonization/fermionization and
the Schwinger model (QED$_{2}$).

The Lenz-Ising (L-I) model was proposed in 1920 by Wihelm Lenz \cite{Lenz} as
the simplest discrete statistical mechanics model with a chance to go beyond
the P. Weiss phenomenological Ansatz involving long range forces and instead
explain ferromagnetism in terms of non-magnetic short range interactions. Its
one-dimensional version was solved 4 years later by his student Ernst Ising.
In his 1925 university of Hamburg thesis, Ising \cite{Lenz} not only showed
that his chain solution could not account for ferromagnetism, but he also
proposed some (as it turned out much later) not entirely correct intuitive
arguments to the extend that this situation prevails to the higher dimensional
lattice version. His advisor Lenz as well as Pauli (at that time Lenz's
assistant) accepted these reasonings and as a result there was considerable
disappointment among the three which resulted in Ising's decision (despite
Lenz's high praise of Ising's thesis) to look for a career outside of
research. For many years a reference by Heisenberg \cite{testing} (to promote
his own proposal as an improved description of ferromagnetism) to Ising's
negative result was the only citation; the situation begun to change when
Peierls \cite{testing} drew attention to ``Ising's solution'' and the results
of Kramers and Wannier \cite{testing} cast doubts on Ising's intuitive
arguments beyond the chain solution. The rest of this fascinating episode i.e.
Lars Onsager's rigorous two-dimensional solution exhibiting ferromagnetic
phase transition, Brucia Kaufman's simplification which led to conceptual and
mathematical enrichments (as well as later contributions by many other
illustrious personalities) hopefully remains a well-known part of mathematical
physics history even beyond my own generation.

This work marks the beginning of applying rigorous mathematical physics
methods to solvable two-dimensional models as the ultimate control of
intuitive arguments in statistical mechanics and quantum field theory. The L-I
model continued to play an important role in the shaping of ideas about
universality classes of critical behavior; in the hands of Leo Kadanoff it
became the key for the development of the concepts about order/disorder
variables (The microscopic version of the famous Kramers-Wannier duality) and
also of the operator product expansions which he proposed as a concrete
counterpart to the more general field theoretic setting of Ken Wilson. Its
massless version (and the related so-called Coulomb gas representation) became
a role model in the setting of the BPZ minimal chiral models (Belavin,
Polyakov and Zamolodchikov 1984) and it remained up to date the only model
with non-abelian braid group (plektonic) statistics for which the n-point
correlators can be written down explicitly in terms of elementary functions
\cite{testing}. Chiral theories confirmed the pivotal role of ``exotic''
statistics \cite{testing} in low dimensional QFT by exposing the appearance of
braid group statistics as a novel manifestation of Einstein causality
\cite{testing}.

Another conceptually rich model which lay dormant for almost two decades as
the result of a misleading speculative higher dimensional generalization by
its protagonist is the bosonization/fermionization model first proposed by
Pascual Jordan \cite{Jor1}. This model establishes a certain equivalence
between massless two-dimensional Fermions and Bosons; it is related to
Thirring's massless 4-fermion coupling model and also to Luttinger's
one-dimensional model of an electron gas \cite{testing}. One reason why even
nowadays hardly anybody knows Jordan's contribution is certainly the ambitious
but unfortunate title ``the neutrino theory of light'' under which he
published a series of papers; besides some not entirely justified criticism of
content, the reaction of his contemporaries consisted in a good-humored
carnivalesque Spottlied (mockery song) about its title \cite{testing}. The
massive version of the related Thirring model became the role model of
integrable relativistic QFT and shed additional light on two-dimensional
bosonization \cite{factorizing}.

Both discoveries demonstrate the usefulness of having controllable
low-dimensional models; at the same time their complicated history also
illustrates the danger of rushing to premature ``intuitive'' conclusions about
extensions to higher dimensions. The search for the appropriate higher
dimensional analog of a 2-dimensional observation is an extremely subtle
endeavour. In the aforementioned two historical examples the true physical
message of those models only became clear through hard mathematical work and
profound conceptual analysis by other authors many years after the discovery
of the original model.

A review of the early historical benchmarks of conceptual progress through the
study of solvable two-dimensional models would be incomplete without
mentioning Schwinger's proposed solution \cite{Schwinger} of two-dimensional
quantum electrodynamics, afterwards referred to as the Schwinger model.
Schwinger used this model in order to argue that gauge theories are not
necessarily tied to zero mass vector particles. Some work was necessary
\cite{testing} to unravel its physical content with the result that the
would-be charge of that QED$_{2}$ model was ``screened'' and its apparent
chiral symmetry broken; in other words the model exists only in the so-called
Schwinger-Higgs phase with massive free scalar particles accounting for its
physical content. Another closely related aspect of this model which also
arose in the Lagrangian setting of 4-dimensional gauge theories was that of
the $\theta$-angle parametrizing, an ambiguity in the quantization.

Thanks to its property of being superrenormalizable, the Schwinger model also
served as a useful testing ground for the Euclidean integral formulation in
the presence of Atiyah-Singer zero modes and their role in the Schwinger-Higgs
chiral symmetry breaking \cite{testing}. These classical topological aspects
of the functional integral formulation attracted a lot of attention beginning
in the late 70s and through the 1980s but, as most geometrical aspects of the
Euclidean functional integral representations, their intrinsic physical
significance remained controversial\footnote{Even in those superrenormalizable
2-dim models, where the measure theory underlying Euclidean functional
integration can be mathematically controlled \cite{Gl-Ja}, there is no good
reason why within this measure theoretical setting outside of quasiclassical
approximations topological properties derived from continuity requirements
should assert themselves.}. This is no problem in the operator algebra
approach where no topological or differential geometrical property is imposed
but certain geometric structures (spacetime- and internal- symmetry
properties) are encoded in the causality and spectral principles of observable algebras.

\ A coherent and systematic attempt at a mathematical control of
two-dimensional models came in the wake of Wightman%
\'{}%
s first rigorous programmatic formulation of QFT \cite{testing}. This
formulation stayed close to the ideas underlying the impressive success of
renormalized QED perturbation theory, although it avoided the direct use of
Lagrangian quantization. The early attempts towards a ``constructive QFT''
found their successful realization in two-dimensional QFT (the $P\varphi_{2}$
models \cite{Gl-Ja}). Only in low dimensional theories the presence of Hilbert
space positivity and energy positivity can be reconciled with the kind of mild
short distance singularity behavior (superrenormalizability) which this
functional analytic method requires. For this reason we will focus our main
attention on alternative constructive methods which are free of this
restrictions; they have the additional advantage to reveal more about the
conceptual structure of QFTs beyond the mere assertion of their existence. The
best illustration of the constructive power of these new methods comes from
massless d=1+1 conformal and chiral QFT as well as from massive factorizing
models. Their presentation and that of the conceptual message they contain for
QFT in general will form the backbone of this article.

There are several books and review articles \cite{chiral} on d=1+1 conformal
as well a on massive factorizing models \cite{factorizing}\cite{Olalla}. To
the extend that concepts and mathematical structures are used which permit no
extension to higher dimensions (Kac-Moody algebras, loop groups,
integrability, presence of an infinite number of conservation laws), this line
of approach will not be followed in this report since our primary interest
will be the use of two-dimensional models of QFT as ``theoretical
laboratories'' of general QFT. Our aim is two-fold; on the one hand we intend
to illustrate known principles of general QFT in a mathematically controllable
context and on the other hand we want to identify new concepts whose
adaptation to QFT in d=1+1 lead to their solvability. In emphasizing the
historical side of the problem, I also hope to uphold the awareness of the
unity and historical continuity in QFT in times of rapidly changing fashions.

Although this article deals with problems of mathematical physics, the style
of presentation is more on a narrative side as expected from an encyclopedia
of mathematical physics contribution. I have tried to amend for the lack of
references (which treat conformal and factorizing models under one roof as
part of QFT) by referring at many instances to a broader review article
\cite{testing} in which most of the left out references and additional related
informations can be found. This review was especially written to serve as a
reference which permits me to maintain an equilibrium between the present size
of text and references.

\section{General concepts and their two-dimensional manifestation}

The general framework of QFT, to which the rich world of controllable
two-dimensional models contributes as an important testing ground, exists in
two quite different but nevertheless closely related formulations: the 1956
approach in terms of pointlike covariant fields due to Wightman \cite{St-Wi},
and the more algebraic setting which can be traced back to ideas which Haag
developed shortly after \cite{Haag} and which are based on spacetime-indexed
operator algebras and related concepts which developed over a long period of
time with contributions of many other authors into what is now referred to as
algebraic QFT (AQFT) or simply local quantum physics (LQP). Whereas the
Wightman approach aims directly at the (not necessarily observable) quantum
fields, the operator algebraic setting ($\rightarrow$ (78), \textit{Algebraic
approach to quantum field theory}) is more ambitious. It starts from
physically well-motivated assumptions about the algebraic structure of local
observables and aims at the reconstruction of the full field theory (including
the operators carrying the superselected charges) in the spirit of a local
representation theory of (the assumed structure of the) local observables.
This has the advantage that the somewhat mysterious concept of an inner
symmetry (as opposed to outer (spacetime) symmetry) can be traced back to its
physical roots which is the representation theoretical structure of the local
observable algebra ($\rightarrow$ (88), \textit{Symmetries of lower spacetime
dimensions}). In the standard Lagrangian quantization approach the inner
symmetry is part of the input (multiplicity indices of field components on
which subgroups of U(n) or O(n) act linearly) and hence it is not possible to
problematize this fundamental question. When in low-dimensional spacetime
dimensions the sharp separation (the Coleman-Mandula theorems) of inner versus
outer symmetry becomes blurred as a result of the appearance of braid group
statistics, the standard Lagrangian quantization setting of most of the
textbooks is inappropriate and even the Wightman framework has to be extended.
In that case the algebraic approach is the most appropriate.

The important physical principles which are shared between the Wightman
approach (WA) \cite{St-Wi} and the operator algebra (AQFT) setting \cite{Haag}
are the spacelike locality or Einstein causality (in terms of pointlike fields
or algebras localized in causally disjoint regions) and the existence of
positive energy representations of the Poincar\'{e} group implementing
covariance and the stability of matter.

The observable algebra consists of a family of (weakly closed) operator
algebras $\left\{  \mathcal{A(O)}\right\}  _{\mathcal{O\in K}}$ indexed by a
family of convex causally closed spacetime regions $\mathcal{O}$ (with
$\mathcal{O}^{\prime}$ denoting the spacelike complement and $\mathcal{A}%
^{\prime}$ the von Neumann commutant) which act in one common Hilbert space.
Certain properties cannot be naturally formulated in the pointlike field
setting (vis. Haag duality$\footnote{Haag duality holds for for observable
algebras in the vacuum sector in the sense that any violation can be explained
in terms of a spontaneously broken symmetry; in local theories it always can
be enforced by dualization and the resulting Haag dual algebra has a charge
superselection structure associated with the unbroken subgroup.}$ for convex
regions $\mathcal{A}(\mathcal{O}^{\prime})=\mathcal{A}(\mathcal{O})^{\prime
}),$ but apart from those properties the two formulations are quite close; in
particular for two-dimensional theories there are convincing arguments that
one can pass between the two without imposing additional technical requirements.

The two above requirements are often (depending on what kind of structural
properties one wants to derive) complemented by additional impositions which,
although not carrying the universal weight of principles nevertheless
represent natural assumptions whose violation, even though not prohibited by
the principles, would cause paradigmatic attention and warrants special
explanations. Examples are ``weak additivity'', ``Haag duality'' and ``the
split property''. Weak additivity i.e. the requirement $\vee\mathcal{A(O}%
_{i}\mathcal{)}=\mathcal{A(O)}$ if $\mathcal{O}=\cup\mathcal{O}_{i}$ expresses
the ``global from amalgamating the local'' aspect which is inherent in the
``action in the neighborhood'' property of fields.

Haag duality is the statement that the commutant of observables not only
contains the algebra of the causal complement (Einstein causality) but is even
exhausted by it i.e. $\mathcal{A}(\mathcal{O}^{\prime})=\mathcal{A}%
(\mathcal{O})^{\prime};$ it is deeply connected to the measurement process and
its violation in the vacuum sector for convex causally complete regions
signals spontaneous symmetry breaking in the associated charge-carrying field
algebra \cite{Haag}. It always can be enforced (assuming that the
wedge-localized algebras fulfill (\ref{mod}) below) by symmetry-reducing
extension called Haag-dualization. Its violation for multi-local region
reveals the charge content of the model via charge-anticharge splitting in the
neutral observable algebra \cite{testing}.

The split property for regions $\mathcal{O}_{i}$ separated by a finite
spacelike distance $\mathcal{A}(\mathcal{O}_{1}\cup\mathcal{O}_{2}%
)\simeq\mathcal{A}(\mathcal{O}_{1})\otimes\mathcal{A}(\mathcal{O}_{2})$
(Doplicher-Longo 1984) is a result of the adaptation of the ``finiteness of
phase space cell'' property of QM to QFT (the so-called ``nuclearity
property''). Related to the Haag duality is the local version of the ``time
slice property'' (the QFT counterpart of the classical causal dependency
property) sometimes referred to as ``strong Einstein causality''
$\mathcal{A}(\mathcal{O}^{\prime\prime})=\mathcal{A}(\mathcal{O}%
)^{\prime\prime}$ \cite{testing}$.$

One of the most astonishing achievements of the algebraic approach is the DHR
theory of superselection sectors (Doplicher, Haag and Roberts, 1971) i.e. the
realization that the structure of charged (non-vacuum) representations (with
the superposition principle being valid only within one representation) and
the spacetime properties of the fields which are the carriers of these
generalized charges, including their spacelike commutation relation which lead
to the particle statistics and also to their internal symmetry properties, are
already encoded in the structure of the Einstein causal observable algebra
($\rightarrow$\ (87) Symmetries in quantum field theory: algebraic aspects).
The intuitive basis of this remarkable result (whose prerequisite is locality)
is that one can generate charged sectors by spatially separating charges in
the vacuum (neutral) sector and disposing of the unwanted charges at spatial
infinity \cite{Haag}.

An important concept which especially in d=1+1 has considerable constructive
clout is ``modular localization''. It is a consequence of the above algebraic
setting if either the net of algebras have pointlike field generators, or if
the one-particle masses are separated by spectral gaps so that the formalism
of time dependent scattering can be applied \cite{testing}; in conformal
theories this property holds automatically in all spacetime dimensions. It
rests on the basic observation ($\rightarrow$ (19) Tomita-Takesaki modular
theory) that a \textit{standard pair} ($\mathcal{A},\Omega$) of a von Neumann
operator algebra and a vector\footnote{Standardness means that the operator
algebra of the pair ($\mathcal{A},\Omega$) act cyclic and separating on the
vector $\Omega.$} gives rise to a Tomita operator S through its star-operation
whose polar decomposition yield two modular objects, a 1-parametric subgroup
$\Delta^{it}$ of the unitary group of operators in Hilbert space whose
Ad-action defines the modular automorphism of ($\mathcal{A},\Omega$) whereas
the angular part $J$ is the modular conjugation which maps $\mathcal{A}$ into
its commutant $\mathcal{A}^{\prime}$
\begin{align}
&  SA\Omega=A^{\ast}\Omega,\;\;S=J\Delta^{\frac{1}{2}}\label{mod}\\
J_{W}  &  =U(j_{W})=S_{scat}J_{0},\;\Delta_{W}^{it}=U(\Lambda_{W}(2\pi
t))\nonumber\\
&  \sigma_{W}(t):=Ad\Delta_{W}^{it}\nonumber
\end{align}
The standardness assumption is always satisfied for any field theoretic pair
($A(\mathcal{O}),\Omega$) of a $O$-localized algebra and the vacuum state (as
long as $\mathcal{O}$ has a nontrivial causal disjoint $\mathcal{O}^{\prime}$)
but it is only for the wedge region $W$ that the modular objects have a
physical interpretation in terms of the global symmetry group of the vacuum as
specified in the second line (\ref{mod}); the modular unitary represents the
$W$-associated boost $\Lambda_{W}(\chi)$ and the modular conjugation
implements the TCP-like reflection along the edge of the wedge
(Bisognano-Wichmann 1976). The third line is the definition of the modular
group. Its usefulness results from the fact that it does not depend on the
state vector $\Omega$ but only on the state $\omega(\cdot)=\left(
\Omega,\cdot\Omega\right)  $ which it induces, as well as the fact that the
modular group $\sigma^{(\eta)}(t)$ associated with a different state $\eta(.)$
on is unitarily equivalent to $\sigma^{(\omega)}(t)$ with a unitary $u(t)$
which fulfills the Connes cocycle property. The importance of this theory for
local quantum physics results from the fact that it leads to concept of
modular localization, a new intrinsic new scenario for field theoretic
constructions which is different from the Lagrangian quantization schemes
\cite{testing}.

A special feature of d=1+1 Minkowski spacetime is the disconnectedness of the
right/left spacelike region leading to a right-left ordering structure. So in
addition to the Lorentz invariant timelike ordering $x\prec y$ (x earlier than
y, which is independent of spacetime dimensions), there is an invariant
spacelike ordering $x<y$ (x to the left of y) in d=1+1 which opens the
possibility of more general Lorentz-invariant spacelike commutation relation
than those implemented by Bose/Fermi fields e.i. of fields with a spacelike
braid group commutation structure. The appearance of such exotic statistics
fields is not compatible with their Fourier transforms being
creation/annihilation operators for Wigner particles; rather the state vectors
which they generate from the vacuum contain in addition to the one-particle
contribution a vacuum polarization cloud \cite{testing}. This close connection
between new kinematic possibilities and interactions is one of the reasons
why, different from higher dimensions where interactions are prescribed by the
recipe of local couplings of free fields, low dimensional QFT offers a more
intrinsic access to the central issue of interactions.

Although the operator-algebraic formulation is well-suited to such a more
intrinsic approach, this does not mean that pointlike covariant fields have
become less useful. They only changed their role; instead of mediating between
classical and quantum field theory in the process of (canonical or functional
integral) quantization, they now are universal generators of all local
algebras and hence also of all modular objects $\Delta_{\mathcal{O}}%
^{it},J_{\mathcal{O}}$ which taken together form an infinite dimensional
noncommutative unitary group in the Hilbert space. This universal group
generated by the modular unitaries contains in particular the global spacetime
symmetry group of the vacuum (Poincar\'{e} transformations, conformal
transformations) as well as ``partial diffeomorphisms'' (section 8).

\section{Boson/Fermion equivalence and superselection theory in a special model}

The simplest and oldest but conceptually still rich model is obtained, as
first proposed by Pascual Jordan \cite{Jor1}, by using a 2-dim. massless Dirac
current and showing that it may be expressed in terms of scalar canonical Bose
creation/annihilation operators
\begin{equation}
j_{\mu}=:\overline{\psi}\gamma_{\mu}\psi:=\partial_{\mu}\phi,\,\phi
:=\int_{-\infty}^{+\infty}\{e^{ipx}a^{\ast}(p)+h.c.\}\frac{dp}{2\left|
p\right|  }%
\end{equation}
Although the potential $\phi(x)$ of the current as a result of its infrared
divergence is not a field in the standard sense of an operator-valued
distribution in the Fock space of the $a(p)^{\#}\footnote{It becomes an
operator after smearing with test functions whose Fourier transform vanishes
at p=0.},$ the formal exponential defined as the zero mass limit of a
well-defined exponential free massive field
\begin{equation}
:e^{i\alpha\phi(x)}:=lim_{m\rightarrow0}m^{\frac{\alpha^{2}}{2}}%
:e^{i\alpha\phi_{m}(x)}: \label{mass}%
\end{equation}
turns out to be a bona fide quantum field in a larger Hilbert space (which
extends the Fock space generated from applying currents to the vacuum). The
power in front is determined by the requirement that all Wightman functions
(computed with the help of free field Wick combinatorics) stay finite in this
massless limit; the necessary and sufficient condition for this is the charge
conservation rule%

\begin{equation}
\left\langle \prod_{i}:e^{i\alpha_{i}\phi(x)}:\right\rangle =\left\{
\begin{array}
[c]{c}%
\prod_{i<j}\left(  \frac{-1}{\left(  \xi_{+ij}\right)  _{\varepsilon}\left(
\xi_{-ij}\right)  _{\varepsilon}}\right)  ^{\frac{1}{2}\alpha_{i}\alpha_{j}%
},\,\sum\alpha_{i}=0\\
0,\,\,\,\,\,\,otherwise
\end{array}
\right.  \label{conservation}%
\end{equation}
where the resulting correlation function has been factored in terms of
lightray coordinates $\xi_{\pm ij}=x_{\pm i}-x_{\pm j},$ $x_{\pm}=t\pm x$ and
the $\varepsilon$-prescription stands for taking the standard Wightman
$t\rightarrow t+i\varepsilon,$ $lim_{\varepsilon\rightarrow0}$ boundary value
which insures the positive energy condition. The additional presence in the
vacuum expectation values of an arbitrary polynomial in the current $\prod
_{i}j_{\mu}{}_{i}(y_{i})$ does not change the argument leading to the charge
conservation law \ref{conservation}. The finiteness of the limit insures that
the resulting zero mass limiting theory is a bona fide quantum field theory
i.e. its system of Wightman functions which permits the construction of an
operator theory in a Hilbert space with a distinguished vacuum vector. There
exists another very intuitive and physically more intrinsic method in which
one stays in the zero mass setting and obtains the charged sectors by
splitting neutral operators as $expij(f)$ belonging to the vacuum sector and
``dumping the unwanted compensating charge behind the moon'' \cite{Haag} by
taking suitable sequences of test function and adjusting normalizations appropriately.

The factorization into lightray components (\ref{conservation}) shows that the
exponential charge-carrying operators inherit this factorization into two
independent chiral components :$expi\alpha\phi(x)$: =:$expi\alpha\phi
_{+}(x_{+})$::$expi\alpha\phi_{-}(x_{-})$: each one being invariant under
scaling $\xi\rightarrow\lambda\xi$ if one assigns the scaling dimension
$d=\frac{\alpha^{2}}{2}$ to the chiral exponential field and $d=1$ to the
current. As any Wightman field this is a singular object which only after
smearing with Schwartz test functions yields an (unbounded) operator. But the
above form of the correlation function belongs to a class of distributions
which admits a much larger test function space consisting of smooth functions
which instead of decreasing rapidly only need to be bounded so that they stay
finite on the compactified lightray line $\dot{R}=S^{1.}$ To make this visible
one uses the Cayley transform (now $x$ denotes either $x_{+}$ or $x_{-}$)%
\begin{equation}
z=\frac{1+ix}{1-ix}\in S^{1}%
\end{equation}
This transforms the Schwartz test function into a space of test functions on
$S^{1}$ which have an infinite order zero at $z=-1$ (corresponding to
$x=\pm\infty$) but the rotational transformed fields $j(z),\,$:$expi\alpha
\phi(z)$: permit the smearing with all smooth functions on $S^{1},$ a
characteristic feature of all conformal invariant theories as the present one
turns out to be. There is an additional advantage in the use of this
compactification. Fourier transforming the circular current actually allows
for a quantum mechanical zero mode whose possible non zero eigenvalues
indicate the presence of additional charge sectors beyond the charge zero
vacuum sector. For the exponential field this leads to a quantum mechanical
pre-exponential factor which \textit{automatically} insures the charge
selection rules (in agreement with the non availability of the ``compensating
charge behind the moon'' argument) so that unrestricted (by charge
conservation) Wick contraction rules can be applied. In this approach the
original chiral Dirac Fermion $\psi(x)$ (from which the current was formed as
the :$\bar{\psi}\psi$: composite) re-appears as a charge-carrying exponential
field for $\alpha=1$ and thus illustrates the meaning of
bosonization/fermionization\footnote{It is interesting to note that Jordan's
original treatment \cite{Jor1} of fermionization had such a pre-exponential
quantum mechanical factor.}. Naturally this terminology has to be taken with a
grain of salt in view of the fact that the bosonic current algebra only
generates a superselected subspace into which the charge-carrying exponential
field does not fit. Only in the case of massive 2-dim. QFT Fermions can be
incorporated into a Fock space of Bosons (see last section). At this point it
should however be clear to the reader that the physical content of Jordan's
paper had nothing to do with its misleading title ``neutrino theory of light''
but rather was a special illustration about charge superselection rules in
QFT, long before this general concept was recognized and formalized.

A systematic and rigorous approach consists in solving the problem of positive
energy representation theory for the Weyl algebra\footnote{The Weyl algebra
originated in quantum mechanics around 1927; its use in QFT only appeared
after the cited Jordan paper. By representation we mean here a regular
representation in which the exponentials can be differentiated in order to
obtain (unbounded) smeare current operators.} on the circle (which is the
rigorous operator algebraic formulation of the abelian current algebra). It is
the operator algebra generated by the exponential of a smeared chiral current
(always with real test functions) with the following relation between the
generators
\begin{align}
&  W(f)=e^{ij(f)},\;\,j(f)=\int\frac{dz}{2\pi i}j(z)f(z),\,\left[
j(z),j(z^{\prime})\right]  =-\delta^{\prime}(z-z^{\prime}),\;\\
&  W(f)W(g)=e^{-\frac{1}{2}s(f,g)}W(f+g),\,\,W^{\ast}(f)=W(-f)\nonumber\\
&  \mathcal{A}(S^{1})=alg\left\{  W(f),f\in C_{\infty}(S^{1})\right\}
,\mathcal{A}(I)=alg\left\{  W(f),suppf\subset I\right\}  \nonumber
\end{align}
where $s(.,.)=\int\frac{dz}{2\pi i}f^{\prime}(z)g(z)$ is the symplectic form
which characterizes the Weyl algebra structure and the last line denotes the
unique $C^{\ast}$ algebra generated by the unitary objects $W(f).$ A
particular representation of this algebra is given by assigning the vacuum
state to the generators $\left\langle W(f)\right\rangle _{0}=e^{-\frac{1}%
{2}\left\|  f\right\|  _{0}^{2}},\,\left\|  f\right\|  _{0}^{2}=\sum_{n\geq
1}n\left|  f_{n}\right|  ^{2}$ Starting with the vacuum Hilbert space
representation $\ \mathcal{A}(S^{1})_{0}=\pi_{0}(\mathcal{A}(S^{1}))$ one
easily checks that the formula
\begin{align}
\left\langle W(f)\right\rangle _{\alpha} &  :=e^{i\alpha f_{0}}\left\langle
W(f)\right\rangle _{0}\\
\pi_{\alpha}(W(f)) &  =e^{i\alpha f_{0}}\pi_{0}(W(f))\nonumber
\end{align}
defines a state with positive energy i.e. one whose GNS representation for
$\alpha\neq0$ is unitarily inequivalent to the vacuum representation. Its
incorporation into the vacuum Hilbert space (second line) is part of the DHR
formalism. It is convenient to view this change as the result of an
application of an automorphism $\gamma_{\alpha}$ on the $C^{\ast}$-Weyl
algebra $\mathcal{A}(S^{1})$ which is implemented by a unitary charge
generating operator $\Gamma_{\alpha}$ in a larger (nonseparable) Hilbert space
which contains all charge sectors $H_{\alpha}=\Gamma_{\alpha}H_{0},H_{0}\equiv
H_{vac}=\overline{\mathcal{A}(S^{1})\Omega}$
\begin{equation}
\left\langle W(f)\right\rangle _{\alpha}=\left\langle \gamma_{\alpha
}(W(f))\right\rangle _{0},\text{ }\gamma_{\alpha}(W(f))=\Gamma_{\alpha
}W(f)\Gamma_{\alpha}^{\ast}%
\end{equation}
$\Gamma_{\alpha}\Omega=\Omega_{\alpha}$ describes a state with a rotational
homogeneous charge distribution; arbitrary charge distributions $\rho_{\alpha
}$ of total charge $\alpha$ i.e. $\int\frac{dz}{2\pi i}\rho_{\alpha}=\alpha$
are obtained in the form
\begin{equation}
\psi_{\rho_{\alpha}}^{\zeta}=\eta(\rho_{\alpha})W(\hat{\rho}_{\alpha}^{\zeta
})\Gamma_{\alpha}\label{form}%
\end{equation}
where $\eta(\rho_{\alpha})$ is a numerical phase factor and the net effect of
the Weyl operator is to change the rotational homogeneous charge distribution
into $\rho_{\alpha}.$ The necessary charge-neutral compensating function
$\hat{\rho}_{\alpha}^{\zeta}$ in the Weyl cocycle $W(\hat{\rho}_{\alpha
}^{\zeta})$ is uniquely determined in terms of $\rho_{\alpha}$ up to the
choice of one point $\zeta\in S^{1}$(the determining equation involves the
$lnz$ function which needs the specification of a branch cut \cite{testing}).
From this formula one derives the commutation relations $\psi_{\rho_{\alpha}%
}^{\zeta}\psi_{\rho_{\beta}}^{\zeta}=e^{\pm i\pi\alpha\beta}\psi_{\rho_{\beta
}}^{\zeta}\psi_{\rho_{\alpha}}^{\zeta}$ for spacelike separations of the
$\rho$ supports; hence these fields are relatively local (bosonic) for
$\alpha\beta=2\mathbb{Z.}$ In particular if only one type of charge is
present, the generating charge is $\alpha_{gen}=\sqrt{2N}\mathbb{\ }$and the
composite charges are multiples i.e. $\alpha_{gen}\mathbb{Z.}$ This locality
condition providing bosonic commutation relations does not yet insure the
$\zeta$-independence. Since the equation which controls the $\zeta$-change
turns out to be%
\begin{equation}
\psi_{\rho_{\alpha}}^{\zeta_{1}}\left(  \psi_{\rho_{\alpha}}^{\zeta_{2}%
}\right)  ^{\ast}=e^{\pm i\pi\alpha\beta}e^{2\pi iQ\alpha}%
\end{equation}
one achieves $\zeta$-independence by restricting the Hilbert space charges to
be ``dual'' to that of the operators i.e. $Q=\left\{  \frac{1}{\sqrt{2N}%
}\mathbb{Z}\right\}  .$ The localized $\psi_{\rho_{\alpha}}^{\zeta_{1}}$
operators acting on the restricted separable Hilbert space $H_{res}$ generate
a $\zeta$- independent extended observable algebra $\mathcal{A}_{N}(S^{1})$
\cite{testing} and it is not difficult to see that its representation in
$H_{res}$ is reducible and that it decomposes into $2N$ charge sectors
$\left\{  \frac{1}{\sqrt{2N}}n\text{, }n=0,1,..N-1\right\}  .$ Hence the
process of extension has led to a charge quantization with a finite
(``rational'') number of charges relative to the new observable algebra which
is neutral in the new charge counting $\frac{1}{\alpha_{gen}}\mathbb{Z}%
/\alpha_{gen}\mathbb{Z}=\mathbb{Z}/\alpha_{gen}^{2}=\mathbb{Z}_{2N}$. The
charge-carrying fields in the new setting are also of the above form
(\ref{form}), but now the generating field carries the charge $\int\frac
{dz}{2\pi i}\rho_{gen}=Q_{gen}$ which is a $\frac{1}{2N}$ fraction of the old
$\alpha_{gen}.$ Their commutation relations for disjoint charge supports are
``braidal'' (or better ``plektonic''\footnote{In the abelian case like the
present the terminology ``anyonic'' enjoys widespread popularity: but in the
present context the ``any'' does not go well with charge quantization.} which
is more on par with bosonic/fermionic). These objects considered as operators
localized on $S^{1}$ do depend on the cut $\zeta,$ but using an appropriate
finite covering of $S^{1}$ this dependence is removed \cite{testing}. So the
field algebra $\mathcal{F}_{\mathbb{Z}_{2N}}$ generated by the charge carrying
fields (as opposed to the bosonic observable algebra $\mathcal{A}_{N}$) has
its unique localization structure on a finite covering of $S^{1}.$ $\,$An
equivalent description which gets rid of $\zeta$ consists in dealing with
operator-valued sections on $S^{1}.$ The extension $\mathcal{A}\rightarrow
\mathcal{A}_{N},$ which renders the Hilbert space separable and quantizes the
charges, seems to be characteristic for abelian current algebra, in all other
models which have been constructed up to now the number of sectors is at least
denumerable and in the more interesting ones even finite (rational models). An
extension is called maximal if there exists no further extension which
maintains the bosonic commutation relation. For the case at hand this would
require the presence of another generating field of the same kind as above
which belongs to an integer $N^{\prime}$ is relatively local to the first one.
This is only possible if $N$ is divisible by a square.

In passing it is interesting to mention a somewhat unexpected relation between
the Schwinger model, whose charges are \textit{screened}, and the Jordan
model. Since the Lagrangian formulation of the Schwinger model is a gauge
theory, the analog of the 4-dim. \textit{asymptotic freedom} wisdom would
suggest the possibility of \textit{charge liberation} in the short distance
limit of this model. This seems to contradict the statement that the intrinsic
content of the Schwinger model (QED$_{2}$ with massless Fermions) (after
removing a classical degree of freedom\footnote{In its original gauge
theoretical form the Schwinger model has an infinite vacuum degeneracy. The
removal of this degeneracy (restoration of the cluster property) with the help
of the ``$\theta$-angle formalism'' leaves a massive free Bose field (the
Schwinger-Higgs mechanism). As expected in d=1+1 the model only possesses this
phase.}) is the QFT of a free massive Bose field and such a simple free field
is at first sight not expected to contain subtle informations about asymptotic
charge liberation. Well, as we have seen above, the massless limit really does
have liberated charges and the short distance limit of the massive free field
is the massless model \cite{testing}.

As a result of the peculiar bosonization/fermionization aspect of the zero
mass limit of the derivative of the massive free field, Jordan's model is also
closely related to the massless Thirring model (and the related Luttinger
model for an interacting one-dimensional electron gas) whose massive version
is in the class of factorizing models (see later section)\footnote{Another
structural consequence of this aspect leads to Coleman's theorem
\cite{testing} which connects the Mermin-Wagner no-go theorem for
two-dimensional spontaneous continuous symmetry-breaking with these zero mass
peculiarities.}. The Thirring model is a special case in a vast class of
``generalized'' multi-coupling multi-component Thirring models i.e. models
with 4-Fermion interactions. Under this name they were studied in the early
70s \cite{testing} with the aim to identify massless subtheories for which the
currents form chiral current algebras.

The counterpart of the potential of the conserved Dirac current in the massive
Thirring model is the Sine-Gordon field, i.e. a composite field which in the
attractive regime of the Thirring coupling again obeys the so-called
Sine-Gordon equation of motion. Coleman gave a supportive argument
\cite{testing} but some fine points about the range of its validity in terms
of the coupling strength remained open\footnote{It was noticed that the
current potential of the free massive Dirac Fermion (g=0) does not obey the
Sine-Gordon equation \cite{testing}.}. A rigorous confirmation of these facts
was recently given in the bootstrap-formfactor setting \cite{testing}. Massive
models which have a continuous or discrete internal symmetry have ``disorder''
fields which implement a ``half-space'' symmetry on the charge-carrying field
(acting as the identity in the other half axis) and together with the basic
pointlike field form composites with have exotic commutation relations (see
last subsection).

\section{The conformal setting, structural results}

Chiral theories play a special role within the setting of conformal quantum
fields. General conformal theories have observable algebras which live on
compactified Minkowski space (S$^{1}$ in the case of chiral models) and
fulfill the Huygens principle, which in an even number of spacetime dimension
means that the commutator is only nonvanishing for lightlike separation of the
fields. The fact that this classical wisdom breaks down for non-observable
conformal fields (e.g. the massless Thirring field) was noticed at the
beginning of the 70s and considered paradoxical at that time
(``reverberation'' in the timelike (Huygens) region). Its resolution around
1974/75 confirmed that such fields are genuine conformal covariant objects but
that some fine points about their causality needed to be addressed. The upshot
was the proposal of two different but basically equivalent concepts about
\textit{globally} causal fields. They are connected by the following global
decomposition formula%

\begin{equation}
A(x_{cov})=\sum A_{\alpha,\beta}(x),\,\,A_{\alpha,\beta}(x)=P_{\alpha
}A(x)P_{\beta},\,\,Z=\sum e^{id_{\alpha}}P_{\alpha} \label{dec}%
\end{equation}

On the left hand side the spacetime point of the field is a point on the
universal covering of the conformal compactified Minkowski space. These are
fields (Luescher and Mack,1976) \cite{testing} which ``live'' in the sense of
quantum (modular) localization on the universal covering spacetime (or on a
finite covering, depending on the ``rationality'' of the model) and fulfill
the global causality condition previously discovered by I. Segal
\cite{testing}. They are generally highly reducible with respect to the center
of the covering group. The family of fields on the right hand side on the
other hand are fields which were introduced (Schroer and Swieca, 1974) with
the aim to have objects which live on the projection $x(x_{cov})$ i.e. on the
spacetime of the physics laboratory instead of the ``hells and heavens'' of
the covering \cite{testing}. They are operator-distributional valued
\textit{sections} in the compactification of ordinary Minkowski spacetime. The
connection is given by the above decomposition formula into irreducible
conformal blocks with respect to the center $\mathbf{Z}$ of the noncompact
covering group $\widetilde{SO(2,n)}$ where $\alpha,\beta$ are labels for the
eigenspaces of the generating unitary $Z$ of the abelian center $\mathbf{Z.}$
The decomposition (\ref{dec}) is minimal in the sense that in general there
generally will be a refinement due to the presence of additional charge
superselection rules (and internal group symmetries). The component fields are
not Wightman fields since they annihilate the vacuum if the right hand
projection differs from $P_{0}=P_{vac}.$

Note that the Huygens (timelike) region in Minkowski spacetime has an timelike
ordering structure $x\prec y$ or $x\succ y$ (earlier, later). In d=1+1 the
topology allows in addition a spacelike left-right ordering $x\lessgtr y.$ In
fact it is precisely the presence of this two orderings in conjunction with
the factorization of the vacuum symmetry group $\widetilde{SO(2,2)}%
\simeq\widetilde{PSL(2R)_{l}}\otimes\widetilde{PSL(2,R)_{r}}$ in particular
$\mathbf{Z\;}\mathbb{=}\mathbf{Z}_{l}\otimes\mathbf{Z}_{r},$ which is at the
root of a significant simplification. This situation suggested a tensor
factorization into chiral components and led to an extremely rich and
successful construction program of two-dimensional conformal QFT as a two-step
process: the classification of chiral observable algebras on the lightray and
the amalgamation of left-right chiral theories to 2-dimensional local
conformal QFT. The action on the circular coordinates $z$ is through
fractional $SU(1,1)$ transformations $g(z)=\frac{\alpha z+\beta}{\bar{\beta
}z+\bar{\alpha}}$ whereas the covering group acts on the Mack-Luescher
covering coordinates.

The presence of an ordering structure permits the appearance of more general
commutation relations for the above $A_{\alpha\beta}$ component fields namely%

\begin{equation}
A_{\alpha,\beta}(x)B_{\beta,\gamma}(y)=\sum_{\beta^{\prime}}R_{\beta
,\beta^{\prime}}^{\alpha,\gamma}B_{\alpha,\beta^{\prime}}(y)A_{\beta^{\prime
},\gamma}(x),\,\,x>y
\end{equation}
with numerical $R-$coefficients which, as a result of associativity and
relative commutativity with respect to observable fields have to obey certain
structure relations; in this way Artin braid relations emerge as a new
manifestation of the Einstein causality principle for observables in
low-dimensional QFT (Rehren and Schroer, 1989) \cite{testing}. Indeed the DHR
method to interpret charged fields as charge-superselection carriers (tied by
local representation theory to the bosonic local structure of observable
algebras) leads precisely to such a \textit{plektonic }statistics structure
(Fredenhagen, Rehren and Schroer, 1992, Froehlich and Gabbiani 1993) for
systems in low spacetime dimension ($\rightarrow$(88) Symmetries of lower
spacetime dimensions). With an appropriately formulated adjustment to
observables fulfilling the Huygens commutativity, this plektonic structure
(but now disconnected from particle/field statistics) is also a possible
manifestation of causality for the higher dimensional timelike structure
\cite{testing}.

Although the above presentation may have created the impression that there is
a straight line from the decomposition theory of the early 70s to the
construction of interesting models, the is not quite the way history unfolded.
The only examples known up to the appearance of the seminal BPZ work (Belavin,
Polyakov and Zamolodchikov, 1984) were the abelian current models of the
previous section which furnish a rather poor man's illustration of the
richness of the decomposition theory. The floodgates of conformal QFT were
only opened after the BPZ discovery of ``minimal models'' which was preceded
by the observation (Friedan, Qiu and Shenker 1984) that the algebra of the
stress-energy tensor came with a new representation structure which was not
compatible with an underlying internal group symmetry ($\rightarrow$(87)
Symmetries and conservation laws).

The importance of the stress-energy tensor in two-dimensional QFT in its role
as a generator of a new infinite-dimensional Lie algebra was already
recognized soon after Virasoro's extraction of part of this algebra from
Veneziano's dual model, but the first field theortic derivations were limited
to the stress-energy tensor of a free massless Dirac field. There was however
another more traditional line of structural arguments which originated in
Wightman's formulation of QFT ($\rightarrow(317)$ Axiomatic quantum field
theory) wherein one was trying to go beyond free fields by staying close to
free field algebraic structures. The first attempt beyond the generalized free
field commutation relations was O. W. Greenberg's proposal to investigate
fields with a Lie-type of spacetime commutation relations i.e. a set of fields
$A_{i}(x)$ fulfilling the ``Lie relation'' (Greenberg, 1961)%
\begin{equation}
\left[  A_{i}(x),A_{j}(y)\right]  =c-number+\int C_{ij}^{k}(x,y,z)A_{k}%
(z)d^{n}z\label{Lie}%
\end{equation}
The non-abelian chiral current algebras at the beginning of the 70s gave some
obvious illustrations of this structure, but the more interesting case was
that of the generic chiral stress-energy tensor \cite{testing}. Later it was
shown (Baumann 1976) that there can be no (scalar) Lie fields in higher
spacetime dimensions \cite{testing} i.e. $C_{ij}^{k}(x,y,z)\equiv0$. Examples
of conformal Lie fields are the current algebras and some of the so-called
W-algebras (generalizations of the stress-energy algebra). We will see in
section 7 that solvable (factorizable) massive two-dimensional theories are
characterized by a different algebraic structure.

\section{Chiral fields and 2-dimensional conformal models}

Let us start with a family which generalizes the abelian model of the previous
section. Instead of a one-component abelian current we now take n independent
copies. The resulting multi-component Weyl algebra has the previous form
except that the current is n-component and the real function space underlying
the Weyl algebra consists of functions with values in an n-component real
vector space $f\in LV$ with the standard Euclidean inner product denoted by
$(,)$. The local extension now leads to $\left(  \alpha,\beta\right)
\in2\mathbb{Z}$ i.e. an even integer lattice $\mathcal{L}$ in $V,$ whereas the
restricted Hilbert subspace $H_{\mathcal{L}^{\ast}}$ which ensures $\zeta
$-independence is associated with the dual lattice $L^{\ast}:$ $\left(
\lambda_{i},\alpha_{k}\right)  =\delta_{ik}$ which contains $\mathcal{L}.$ The
resulting superselection structure (i.e. the $Q-$spectrum) corresponds to the
finite factor group $\mathcal{L}^{\ast}/\mathcal{L}.$ For selfdual lattices
$\mathcal{L}^{\ast}=\mathcal{L}$ (which only can occur if dimV is a multiple
of 8) the resulting observable algebra has only the vacuum sector; the most
famous case is the Leech lattice $\Lambda_{24}$ in $dimV=24,$ also called the
``moonshine'' model. The observation that the root lattices of the Lie
algebras of type $A,B$ or $E$ (example. $su(n)$ corresponding to $A_{n-1}$)
also appear among the even integral lattices suggests that the nonabelian
current algebras associated to those Lie algebras can also be implemented.
This turns out to be indeed true as far as the level 1 representations are
concerned which brings us to the second family: the nonabelian current
algebras of level $k$ associated to those Lie algebras; they are characterized
by the commutation relation
\begin{equation}
\left[  J_{\alpha}(z),J_{\beta}(z^{\prime})\right]  =if_{\alpha\beta}^{\gamma
}j_{\gamma}(z)\delta(z-z^{\prime})-\frac{1}{2}kg_{\alpha\beta}\delta^{\prime
}(z-z^{\prime})\label{current}%
\end{equation}
where $f_{\alpha\beta}^{\gamma}$ are the structure constants of the underlying
Lie algebra, $g$ their Cartan-Killing form and $k$, the level of the algebra,
must be an integer in order that the current algebra can be globalized to a
loop group algebra. The Fourier decomposition of the current leads to the so
called affine Lie algebras, a special family of Kac-Moody algebras. For k=1
these currents can be constructed as bilinears in terms of multi-component
chiral Dirac field; there exists also the mentioned possibility to obtain them
by constructing their maximal Cartan currents within the above abelian setting
and representing the remaining non-diagonal currents as certain
charge-carrying (``vertex'' algebra) operators. Level $k$ algebras can be
constructed from reducing tensor products of $k$ level one currents or
directly via the representation theory of infinite-dimensional affine Lie
algebras\footnote{The global exponentiated algebras (the analogs to the Weyl
algebra) are called loop group algebras.}. Either way one finds that e.g. the
$SU(2)$ current algebra of level $k$ has (together with the vacuum sector)
$k+1$ sectors (inequivalent representations). The different sectors are
already distinguished by the structure of their ground states of the conformal
Hamiltonian $L_{0}.$ Although the computation of higher point correlation
functions for $k>1,$ there is no problem in securing the existence of the
algebraic nets which define these chiral models as well as their $k+1$
representation sectors and to identify their generating charge-carrying fields
(primary fields) including their R-matrices appearing in their plektonic
commutation relations. It is customary to use the notation $SU(2)_{k}$ for the
abstract operator algebras associated with the current generators
(\ref{current}) and we will denote their $k+1$ equivalence classes of
representations by $\mathcal{A}_{SU(2)_{k},n},$ $n=0,...k,$ whereas
representations of current algebras for higher rank groups require a more
complicated labeling (in terms of Weyl chambers).

The third family of models are the so-called minimal models which are
associated with the Lie-field commutation structure of the chiral
stress-energy tensor which results from the chiral decomposition of a
conformally covariant 2-dimensional stress-energy tensor
\begin{equation}
\left[  T(z),T(z^{\prime})\right]  =i(T(z)+T(z^{\prime}))\delta^{\prime
}(z-z^{\prime})+\frac{ic}{24\pi}\delta^{\prime\prime\prime}(z-z^{\prime})
\end{equation}
whose Fourier decomposition yields the Witt-Virasoro algebra i.e. a central
extension\footnote{The presence of the central term in the context of QFT (the
analog of the Schwinger term) was noticed later, however the terminology
Witt-Virasoro algebra in the physics literature came to mean the Lie algebra
of diffeomorphisms of the circle including the central extension.} of the Lie
algebra of the $Diff(S^{1})$. The first two coefficients are determined by the
physical role of $T(z)$ as the generating field density for the Lie algebra of
the Poincar\'{e} group whereas the central extension parameter $c>0$
(positivity of the two-point function) for the connection with the generation
of the Moebius transformations and the undetermined parameter $c>0$ (the
central extension parameter) is easily identified with the strength of the
two-point function. Although the structure of the T-correlation functions
resembles that of free fields (in the sense that is a algebraically computable
unique set of correlation functions once one has specified the two-point
function), the realization that $c$ is subject to a discrete quantization if
$c<1$ came as a surprise. As already mentioned, the observation that the
superselection sectors (the positive energy representation structure) of this
algebra did not at all follow the logic of a representation theory of an inner
symmetry group generated a lot of attention and stimulated a flurry of
publications on symmetry concepts beyond groups (quantum groups). A concept of
fundamental importance is the DHR theory of localized endomorphisms of
operator algebras and the concept of operator algebraic inclusions in
particular inclusions with conditional expectations (V. Jones inclusions)

The $SU(2)_{k}$ current coset construction (Goddard, Kent and Olive 1986)
revealed that the proof of existence and the actual construction of the
minimal models is related to that of the $SU(2)_{k}$ current algebras.
Constructing chiral models does not necessarily mean the explicit
determination of its n-point Wightman functions of their generating fields
(which for most chiral models remains a prohibitively complicated) but rather
a proof of their existence by demonstrating that these models are obtained
from free fields by a series of computational complicated but mathematically
controlled operator-algebraic steps as: reduction of tensor products,
formation of orbifolds under group actions, coset constructions and a special
kind of extensions. The generating fields of the models are nontrivial in the
sense of not obeying free field equations (i.e. not being ``on-shell''). The
cases where one can write down explicit n-point functions of generating fields
are very rare; in the case of the minimal family this is only possible for the
Ising model \cite{testing}.

To show the power of inclusion theory for the determination of the charge
content of theory let us look at a simple illustration in the context of the
above multi-component abelian current algebra. The vacuum representation of
the corresponding Weyl algebra is generated from smooth $V$-valued functions
on the circle modulo constant functions (i.e. functions with vanishing total
integral) $f\in LV_{0}$. These functions equipped with the aforementioned
complex structure and scalar product yield a Hilbert space$.$ The
$I$-localized subalgebra is generated by the Weyl image of $I$-supported
functions (class functions whose representing functions are constant in the
complement $I^{\prime}$)
\begin{align}
\mathcal{A}(I)  &  :=alg\left\{  W(f)|\;f\in K(I)\right\} \nonumber\\
K(I)  &  =\left\{  f\in LV_{0}|\,f=const\,\,in\,\,I^{\prime}\right\}
\end{align}
The one-interval Haag duality $\mathcal{A}(I)^{\prime}=\mathcal{A}(I^{\prime
})$ (the commutant algebra equals the algebra localized in the complement) is
simply a consequence of the fact that the symplectic complement $K(I)^{\prime
}$ in terms of $Im(f,g)$ consists of real functions in that space which are
localized in the complement i.e. $K(I)^{\prime}=K(I^{\prime}).$ The answer to
the same question for a double interval $I=I_{1}\cup I_{3}$ (think of the
first and third quadrant on the circle) does not lead to duality but rather to
a genuine inclusion
\begin{align}
&  K(\left(  I_{1}\cup I_{3}\right)  ^{\prime})=K(I_{2}\cup I_{4})\subset
K(I_{1}\cup I_{3})^{\prime}\\
&  \curvearrowright K(I_{1}\cup I_{3})\subset K(\left(  I_{1}\cup
I_{3}\right)  ^{\prime})^{\prime}\nonumber
\end{align}
The meaning of the left hand side is clear, these are functions which are
constant in $I_{1}\cup I_{3}$ with the same constant in the two intervals
whereas the functions on the right hand side are less restrictive in that the
constants can be different. The conversion of real subspaces into von Neumann
algebras by the Weyl functor leads to the algebraic inclusion $\mathcal{A}%
(I_{1}\cup I_{3})\subset\mathcal{A}(\left(  I_{1}\cup I_{3}\right)  ^{\prime
})^{\prime}.$ In physical terms the enlargement results from the fact that
within the charge neutral vacuum algebra a charge split with one charge in
$I_{1}$ and the compensating charge in $I_{2}$ for all values of the
(unquantized) charge occurs. A more realistic picture is obtained if one
allows a charge split is subjected to a charge quantization implemented by a
lattice condition $f(I_{2})-f(I_{4})\in2\pi L$ which relates the two
multi-component constant functions (where $f(I)$ denotes the constant value
$f$ takes in $I$)$.$ As in the previous one-component case the choice of even
lattices corresponds to the local (bosonic) extensions. Although imposing such
a lattice structure destroys the linearity of the $K$, the functions still
define Weyl operators which generated operator algebras $\mathcal{A}_{L}%
(I_{1}\cup I_{2})$\footnote{The linearity structure is recoverd on the level
of the operator algebra.}$.$ But now the inclusion involves the dual lattice
$L^{\ast}$ (which of course contains the original lattice)%
\begin{align*}
&  \mathcal{A}_{L}(I_{1}\cup I_{2})\subset\mathcal{A}_{L^{\ast}}(I_{1}\cup
I_{2})\\
&  ind\left\{  \mathcal{A}_{L}(I_{1}\cup I_{2})\subset\mathcal{A}_{L}(\left(
I_{1}\cup I_{2}\right)  ^{\prime})^{\prime}\right\}  =\left|  G\right| \\
&  \mathcal{A}_{L}(I_{1}\cup I_{2})=inv_{G}\mathcal{A}_{L^{\ast}}(I_{1}\cup
I_{2})
\end{align*}
This time the possible charge splits correspond to the factor group
$G=L^{\ast}/L$ i.e. the number of possibilities is $\left|  G\right|  $ which
measures the relative size of the bigger algebra in terms of the smaller. This
is a special case of the general concept of the so-called Jones index of a an
inclusion which is a numerical measure of its depth. A prerequisite is that
the inclusion permits a conditional expectation which is a generalization of
the averaging under the ``gauge group'' $G$ on $\mathcal{A}_{L^{\ast}}%
(I_{1}\cup I_{2})$ in the third line which identifies the invariant smaller
algebra is the fix point algebra (the invariant part) under the action of $G$.
In fact using the conceptual framework of Jones one can show that the
two-interval inclusion is independent of the position of the disjoint
intervals characterized by the group $G$.

There exists another form of this inclusion which is more suitable for
generalizations. One starts from the quantized charge extended local algebra
$\mathcal{A}_{L}^{ext}\supset\mathcal{A}$ described before in terms of an
integer even lattice $L$ (which lives in the separable Hilbert space
$H_{L^{\ast}})$ as our observable algebra. Again the Haag duality is violated
and converted into an inclusion $\mathcal{A}_{L}^{ext}(I_{1}\cup I_{2}%
)\subset\mathcal{A}_{L}^{ext}(\left(  I_{1}\cup I_{2}\right)  ^{\prime
})^{\prime}$ which turns out to have the same $G=L^{\ast}/L$ charge structure
(it is in fact isomorphic to the previous inclusion). In the general setting
(current algebras, minimal model algebras,...) this double interval inclusion
is particularly interesting if the associated Jones index is finite. One finds
(Kawahigashi-Longo-Mueger 2001) \cite{testing}

\begin{theorem}
A chiral theory with finite Jones index $\mu=ind\left\{  \mathcal{A}(\left(
I_{1}\cup I_{2}\right)  ^{\prime})^{\prime}:\mathcal{A}(I_{1}\cup
I_{2})\right\}  $ for the double interval inclusion (always assuming that
$A(S^{1})$ is strongly additive and split) is a rational theory and the
statistical dimensions $d_{\rho}$ of its charge sectors are related to $\mu$
through the formula
\begin{equation}
\mu=\sum_{\rho}d_{\rho}^{2}%
\end{equation}
\end{theorem}

Instead of presenting more constructed chiral models it may be more
informative to mention some of the algebraic methods by which they are
constructed and explored. The already mentioned DHR theory provides the
conceptual basis for converting the notion of positive energy representation
sectors of the chiral model observable algebras $\mathcal{A}$ (equivalence
classes of unitary representations) into localized endomorphisms $\rho$ of
this algebra. This is an important step because contrary to group
representations which have a natural tensor product composition structure,
representations of operator algebras generally do not come with a natural
composition structure. The DHR endomorphisms theory of $\mathcal{A}$ leads to
fusion laws and an intrinsic notion of generalized statistics (for chiral
theories: plektonic in addition to bosonic/fermionic). The chiral statistics
parameter are complex numbers \cite{Haag} whose phase is related to a
generalized concept of spin via a spin statistics theorem and whose absolute
value (the statistics dimension) generalized the notion of multiplicities of
fields known from the description of inner symmetries in higher dimensional
standard QFTs. The different sectors may be united into one bigger algebra
called the exchange algebra $\mathcal{F}_{red}$ in the chiral context (the
``reduced field bundle'' of DHR) in which every sector occurs by definition
with multiplicity one and the statistics data are encoded into exchange
(commutation) relations of charge-carrying operators or generating fields
(``exchange algebra fields'') \cite{testing}. Even though this algebra is
useful in that all properties concerning fusion and statistics are nicely
encoded, it lacks some cherished properties of standard field theory namely
there is no unique state--field relation i.e. no Reeh-Schlieder
property\footnote{A field $A_{\alpha\beta}$ whose source projection $P_{\beta
}$ does not coalesce with the vacuum projection annihilates the vacuum.}; in
operator algebraic terms, the local algebras are not factors. This poses the
question of how to manufacture from the set of all sectors natural (not
necessarily local) extensions with these desired properties. It was found that
this problem can be characterized in operator algebraic terms by the existence
of so called DHR triples \cite{testing}. In case of rational theories the
number of such extensions is finite and in the aforementioned ``classical''
current algebra- and minimal- models they all have been constructed by this
method\footnote{Thus confirming existing results completing the minimal family
by adding some missing models. }. The same method adapted to the chiral tensor
product structure of d=1+1 conformal observables classifies and constructs all
2-dimensional local (bosonic/fermionic) conformal QFT $\mathcal{B}_{2}$ which
can be associated with the observable chiral input. It turns out that this
approach leads to another of those pivotal numerical matrices which encode
structural properties of QFT: the coupling matrix $Z$%
\begin{align}
\mathcal{A\otimes A} &  \subset\mathcal{B}_{2}\\
\sum_{\rho\sigma}Z_{\rho,\sigma}\rho(\mathcal{A})\otimes\sigma(\mathcal{A}) &
\subset\mathcal{A\otimes A}\nonumber
\end{align}
where the second line is an inclusion solely expressed in terms of observable
algebras from which the desired (isomorphic) inclusion in the first line
follows by a canonical construction, the so-called Jones basic construction.
The numerical matrix $Z$ is an invariant closely related to the so-called
\textit{statistics character matrix} \cite{testing} and in case of rational
models it is even a modular invariant with respect to the modular $SL(2,Z)$
group transformations (which are closely related to the matrix S in section 7).

\section{\bigskip Integrability, the bootstrap-formfactor program}

\textit{Integrability} in QFT and the closely associated
\textit{bootstrap-formfactor} construction of a very rich class of massive
two-dimensional QFTs can be traced back to two observations made during the
60s and 70s ideas. On the one hand there was the time-honored idea to bypass
the ``off-shell'' field theoretic approach to particle physics in favor of a
pure on-shell S-matrix setting which (in particular recommended for strong
interactions), as the result of the elimination of short distances via the
mass-shell restriction would be free of ultraviolet divergencies. This idea
was enriched in the 60s by the crossing property which in turn led to the
bootstrap idea, a highly nonlinear seemingly selfconsistent proposal for the
determination of the S-matrix. However the protagonists of this S-matrix
bootstrap program placed themselves into a totally antagonistic fruitless
position with respect to QFT so that the strong return of QFT in the form of
gauge theory undermined their credibility. On the other hand there were rather
convincing quasiclassical calculations in certain two-dimensional massive QFTs
as e.g. the Sine-Gordon model which indicated that the obtained quasiclassical
mass spectrum is exact and hence suggested that the associated QFTs are
integrable (Dashen-Hasslacher-Neveu 1975) and have no real particle creation.
These provocative observations\footnote{It was believed that the ``nontrivial
elastic scattering implies particle creation'' statement of Aks (Aks, 1963) is
valid also for low-dimensional QFTs.} asked for a structural explanation
beyond quasiclassical approximations, and it became soon clear that the
natural setting for obtaining such mass formulas was that of the fusion of
boundstate poles of unitary crossing-symmetric purely elastic S-matrices;
first in the special context of the Sine-Gordon model (Schroer-Truong-Weisz
1976) and later as a classification program from which factorizing S-matrices
can be determined by solving well-defined equations for the elastic 2-particle
S-matrix (Karowski-Thun-Truong-Weisz 1977). Some equations in this bootstrap
approach resembled mathematical structures which appeared in C. N. Yang's work
on non-relativistic $\delta$-function particle interactions as well as
relations for Boltzmann weights in Baxter's work on solvable lattice models;
hence they were referred to as Yang-Baxter relations. These results suggested
that the old bootstrap idea, once liberated from its ideological dead freight
(in particular from the claim that the bootstrap leads to a unique ``theory of
everything'' (minus gravity)) generates a useful setting for the
classification and construction of factorizing two-dimensional relativistic
S-matrices. Adapting certain known relations between two-particle formfactors
of field operators and the S-matrix to the case at hand (Karowski-Weisz 1978),
and extending this with hindsight to generalized (multiparticle) formfactors,
one arrived at the axiomatized recipes of the bootstrap-formfactor program of
d=1+1 factorizable models (Smirnov 1992). Although this approach can be
formulated within the setting of the LSZ scattering formalism, the use of a
certain algebraic structure (A.B. and Al. B. Zamolodchikov 1979) which in the
simplest version reads%

\begin{align}
Z(\theta)Z^{\ast}(\theta^{\prime})  &  =S^{(2)}(\theta-\theta^{\prime}%
)Z^{\ast}(\theta^{\prime})Z(\theta)+\delta(\theta-\theta^{\prime}%
)\label{Zam}\\
Z(\theta)Z(\theta^{\prime})  &  =S^{(2)}(\theta^{\prime}-\theta)Z(\theta
^{\prime})Z(\theta)\nonumber
\end{align}
which will be referred to as the Z-F algebra (Faddeev added the $\delta$-term)
brought significant simplifications. In the general case the $Z^{\prime}s$ are
vector-valued and the $S^{(2)}$-structure function is
matrix-valued\footnote{The identification of the Z-F structure coefficients
with the elastic two-particle S-matrix $S^{(2)}$ which is prempted by our
notation can be shown to follow from the physical interpretation of the Z-F
structure in terms of localization..}. In that case the associativity of the
Z-F algebra is equivalent to the Yang-Baxter equations. Recently it became
clear that this algebraic relation has a deep physical interpretation; it is
the simplest algebraic structure which can be associated with generators of
nontrivial wedge-localized operator algebras (see next section).

The mentioned quasiclassical integrability observations also led to another
approach which is based on the quantum adaptation of the classical notion of
integrability ($\rightarrow(107)$ Integrable systems: overview). However the
construction of a complete (infinite in field theory) set of conserved
currents with their associated charges in involution is already a detailed and
case-dependent enterprize in the classical setting even before one establishes
the absence of quantum anomalies. Conceptually as well as computationally it
is much simpler to identify the intrinsic meaning of integrability in QFT with
the factorization of its S-matrix or a certain property of wedge-localized
algebras (see next section).

The first step of the bootstrap-formfactor program namely the classification
and construction of model S-matrices follows a combination of two patterns:
prescribing particle multiplets transforming according to group symmetries
and/or specifying structural properties of the particle spectrum. The simplest
illustration for the latter strategy is supplied by the $\mathbb{Z}_{N}$
model. In terms of particle content $\mathbb{Z}_{N}$ demands the
identification of the $N^{th}$ bound state with the antiparticle. Since the
fusion condition for the bound mass $m_{b}^{2}=(p_{1}+p_{2})^{2}=m_{1}%
^{2}+m_{2}^{2}+2m_{1}m_{2}ch(\theta_{1}-\theta_{2})$ is only possible for a
pure imaginary rapidity difference $\theta_{12}=\theta_{1}-\theta_{2}=i\alpha$
(``binding angle''). Hence the binding of two ``elementary''
particles\footnote{The quotation mark is meant to indicate that in contrast to
the Schroedinger QM there is ``nuclear democracy'' on the level of particles.
The inexorable presence of interaction-caused vacuum polarization limits a
fundamental/fused hierarchy to the fusion of charges.} of mass m gives
$m_{2}=m\frac{sin2\alpha}{sin\alpha}$ and more generally of k particles with
$m_{k}=m\frac{sink\alpha}{sin\alpha},$ so that the antiparticle mass condition
$m_{N}=\bar{m}=m$ fixes the binding angle to $\alpha=\frac{2\pi}{N}.$ The
minimal (no additional physical poles) two-particle
S-matrix\footnote{minimal=without so-called CDD poles} in terms of which the
n-particle S-matrix factorizes is therefore%
\begin{equation}
S_{min}^{(2)}=\frac{sin\frac{1}{2}(\theta+\frac{2\pi i}{N})}{sin\frac{1}%
{2}(\theta-\frac{2\pi i}{N})}%
\end{equation}
The SU(N) models as compared with the U(N) model requires a similar
identification of bound states of N-1 particles with an antiparticle. This
S-matrix enters as in the equation for the vacuum to n-particle meromorphic
formfactor of local operators; together with the crossing and the so-called
``kinematical pole equation'' one obtains a recursive infinite system linking
a certain residue with a formfactor involving a lower number of particles. The
solutions of this infinite system form a linear space from which the
formfactors of specific tensor fields can be selected by a process which is
analog but more involved than the specification of a Wick basis of composite
free fields. Although the statistics property of two-dimensional massive
fields is not intrinsic but a matter of choice, it would be natural to realize
e.g. the $\mathbb{Z}_{N}$ fields as $\mathbb{Z}_{N}$-anyons.

Another rich class of factorizing models are the Toda theories of which the
Sine-Gordon and Sinh-Gordon are the simplest cases. For their descriptions the
quasiclassical use of Lagrangians (supported by integrability) turns out to be
of some help in setting up their more involved bootstrap-formfactor construction.

The unexpected appearance of objects with new fundamental (solitonic) charges
(example: the Thirring field as the carrier of a solitonic Sine-Gordon charge)
or the unexpected confinement of charges (example: the $CP(1)$ model as a
confined $SU(2)$ model) turns out to be opposite sides of the same coin and
both cases have realizations in the setting of factorizing models
\cite{testing}.

\section{ Factorizing QFT, PFGs and lightray holography}

There are two recent ideas which place the two-dimensional
bootstrap-formfactor program into a more general setting which permits to
understand its position in the general context of local quantum physics.

Let us restrict our interest to models which fulfill the standard assumption
of LSZ scattering theory (mass gap, asymptotic completeness) and assume for
simplicity just kind of particle. Let $G$ be a (generally unbounded) operator
affiliated with the local algebra $\mathcal{A}(\mathcal{O}).$ We call such $G$
a vacuum-\textbf{p}olarization-\textbf{f}ree \textbf{g}enerator (PFG)
affiliated with $\mathcal{A}(\mathcal{O})$ (denoted as $G\eta\mathcal{A}%
(\mathcal{O})$) if the state vector $G\Omega$ (with $\Omega$ the vacuum) is a
one-particle state without any vacuum polarization admixture \cite{testing}.
PFGs are by definition (unbounded) \textit{on-shell} operators and it is
well-known that the existence of a \textit{subwedge-localized} PFG forces the
theory to be free, i.e. the local algebras in such a situation are generated
by free fields. However, and this is the surprising fact, PFGs and
interactions are compatible in wedge regions. Such localization regions offer
the best compromise between particles localization and vacuum-polarization
favoring field localization\footnote{In any smaller localization region the
interaction-caused vacuum polarizations would only permit field- but not
particle- localization.}. Although modular operator theory guaranties the
existence of wedge generators without vacuum polarization, these PFG have
useful properties in the setting of time-dependent scattering theory only if
they are ``tempered'' (well-defined on a translation invariant domain)
\cite{testing}. The restriction implied by this additional requirement can be
shown to only permit theories with a purely elastic S-matrix and it has been
known for a long time that this is possible only in d=1+1 where such theories
have been investigated since the late 70s within the bootstrap-formfactor
program. In fact on the basis of formfactor properties one can show that the
elastic two-dimensional S-matrices coming from a local QFT are already
described by a two-particle S-matrices \cite{testing}, all the higher elastic
contributions factorize into two-particle contributions and the latter are
classified by solving equations (parametrized in terms of rapidities) which
incorporate unitarity, analyticity and crossing \cite{testing}. The second
surprise is that the Fourier transforms of the wedge-algebra-generating
tempered PFGs are identical to operators introduced at the end of the 70s by
Zamolodchikov (their properties were spelled out in more detail by Faddeev).
Although their usefulness in the bootstrap formfactor program was beyond
doubt, their conceptual position within QFT was not clear since,
notwithstanding their formal similarity to free field creation/annihilation
operators, their physical content is distinctively different from incoming or
outgoing free fields of scattering theory. In the simple case (\ref{Zam}) of
just one interacting particle (without boundstates e.g. the ``sinh-Gordon''
model) the generators of the wedge algebra are of the form%
\begin{equation}
\phi(x)=\frac{1}{\sqrt{2\pi}}\int(e^{ip(\theta)x(\chi)}Z(\theta)+h.c.)d\theta
\label{gen}%
\end{equation}

This defines a covariant field which, although not being pointlike local
maintains some localization; it turns out to be wedge-like local\footnote{The
$x$ continues to comply with the covariant transformation law but it is not a
point of localization i.e. the smearing with wedge supported test functions
$\phi(f)$ does not lead to an improvement in localization if one reduces the
support of $f.$} i.e. it commutes with its ``modular opposite'' $J\phi(x)J$
which generates the causal disjoint algebra $\mathcal{A}(W^{\prime
})=\mathcal{A}(W)^{\prime}.$ This interpretation of the Z-F algebra operators
in terms of localization concepts turns out to be a valuable starting point
for the construction of tighter localized algebras $\mathcal{A}(D)$ associated
with double cone regions $D$ by computing intersections of wedge algebras
whose generating operators turn out to be infinite series in the $Z^{\prime}s$
with coefficient functions which are generalized formfactors. The difficult
problem of demonstrating the existence of a QFT associated with the algebraic
structure (\ref{Zam}) of wedge algebra generators (\ref{gen}) is then encoded
into a nontriviality statement for the double cone intersections
($\mathcal{A}(D)=\mathcal{A}(W_{a})\cap\mathcal{A}(W)^{\prime}\neq
C\mathbf{1,}W_{a}$ translated wedge); the nontriviality of these intersections
for arbitrary small $D$ corresponds to the existence of pointlike generating
fields in the setting of Wightman. The important difference to the standard
perturbative Lagrangian quantization approach is that the computations of
intersections do not require the use of singular correlation functions of
pointlike distributional objects; fields are simply size independent pointlike
generators of local algebras which are convenient for coordinatizing the
resulting local nets of operator algebras but should be avoided in actual
computations where they create the ultraviolet divergence menace.
Unfortunately the standard quantization approach is not able to do this. The
modular wedge localization formalism in terms of PFG generators fulfilling the
Z-F algebra relation for the first time permits to bypass quantization and to
walk without ``classical crutches'' in case of a special interacting class of
factorizable QFT which in their renormalized perturbative quantization
treatment would present all the ultraviolet problems characteristic of the
standard approach. In the absence of interactions Wigner already achieved an
intrinsic formulation in his 1939 one-particle theory. In fact there are good
reasons for viewing the present ideas as an extension of Wigner's approach
into the realm of interactions which as the result of interaction-induced
vacuum polarization delocalize particles and make the introduction of more
convenient carriers of localization unavoidable.

The recognition that the knowledge of the position of a wedge-localized
subalgebra $\mathcal{A}(W)$ with $\mathcal{A}(W)^{\prime}=\mathcal{A}%
(W^{\prime})$ within the full Fock space algebra $B(H)$ and the action of the
representation of the Poincar\'{e} group in $B(H)$ on the $\mathcal{A}(W)$)
determines the full net of algebras $\mathcal{A}(\mathcal{O})$ via
intersections%
\begin{equation}
\mathcal{A}(\mathcal{O}):=\bigcap_{W\supset\mathcal{O}}\mathcal{A}(W)
\end{equation}
is actually independent of spacetime dimensions and factorizability. But only
in d=1+1 within the setting of factorizable models one finds simple PFG
generators for $\mathcal{A}(W)$ which permit the computation of intersections.

With the particle picture outside factorizing theories being made less useful
by de-localization through interaction induces vacuum polarization
\cite{testing} it is encouraging to notice that there is another constructive
idea based on modular inclusion and intersections which does not require the
very restrictive presence of wedge-localized PFGs.. This is the holographic
projection to the lightfront. It maps a massive (non-conformal) QFT to a
(transverse) extended chiral theory on the lightfront $x_{-}=0$ in such a way
that the global original algebra on Minkowski spacetime $\mathcal{A}(M)=B(H)$
and its global holographic lightfront projection $\mathcal{A}(LF)=B(H)$
coalesce\footnote{Conformally invariant theories in d=1+1 are the only
exception to this equality; as a result of the chiral factorization one need
the charateristic data on both lightrays.}, but the local substructure (the
spacetime-indexed net) is radically different, apart from wedge-localized
algebras which are identical to the algebras of their upper lightfront
boundary of $W,$ $\mathcal{A}(W)=\mathcal{A}(LF(W))$. The situation simplifies
considerably in massive two-dimensional theories in which case the transverse
extension is absent and the holographic lightray projection leads to bona-fide
chiral theories (i.e. Moebius-invariant theories whose extendibility to
$Diff(S^{1})$ will be the subject of some remarks in the next section). The
restriction to two-dimensional factorizing theories leads to further
significant simplifications; within this class the holographic projection to
chiral theories has a unique inversion. If the $Diff(S^{1})$ transformations
are present in the holographic projection, they were already present in the
ambient theory. The reason why they are not noticed in the spacetime indexing
of the ambient net is that they acts in a nonclassical ``fuzzy'' manner and
hence escapes the standard symmetry formalism via Noether currents and their
quantization \cite{testing}. The holographic relation has very interesting
consequences for those chiral models (all?) whose observable algebras are in
the holographic image of factorizing massive theories because they can be
characterized in terms of a Z-F algebra structure which in view of its simple
systematics (and the fact that there exists no direct Lie-algebraic structural
characterization for general chiral models) may turn out to provide valuable
additional help in their classifiation.

Besides the rigorous one-to-one relation of factorizing theories to their
chiral holographic projections there is of course also the critical- or
scale-invariant limit which leads to an associated two-dimensional conformal
invariant theory. The idea that the critical universality classes are much
smaller within the setting of factorizing models is the starting point of a
Zamolodchikov%
\'{}%
s successful proposal to approach the classification and identification of
factorizing models via perturbing the better known conformal models by
selecting particular perturbations in terms of chiral operators. Of course
even in the limited factorized setting one cannot expect a one-to one
correspondence since the relation to the conformal massless limits is not just
a simple mass-dressing of fields which already exist on the massless level
(examples show that there are fields in the massive model which vanish if the
massless limit is taken as in section3). However for the consistency of the
Zamolodchikov perturbation scenario one does not have to construct
\textit{all} massive fields directly; it suffices that the ones which
disappear in the massless limit are composites of those who persist. The
universality class division of massive theories by the scale-invariant limit
is conceptually very different from that of the holographic projection; the
latter involves a different encoding of spacetime indexing, but does not
affect the algebraic ``stem cell substrate''\footnote{I find this analogy
quite helpful for a more intrinsic understanding of how QFT processes the
abstract algebraic substrate into various different spacetime-indexed
algebraic nets.} which can be grown into different QFTs by changing the
(curved) space-time indexing (see also next section).

It turns out that the plektonic relations of charged fields and the issue of
statistics of particles loose their physical relevance for two-dimensional
massive models since they can be changed without affecting the physical
content. Instead such notions as order/disorder fields and soliton take their
place \cite{testing}.

In accordance with its historical origin the theory of two-dimensional.
factorizing models may also be is an outgrowth of the quantization of
classical integrable systems ($\rightarrow$(363) \textit{Integrable models in
two dimensions}). But in comparison with the rather involved structure of
integrabilty (existence of sufficiently many commuting conservation laws), the
conceptual setting of factorizing models within the scattering framework
(factorization$\simeq$existence of wedge-localized tempered PFGs) is rather
simple and intrinsic.

There are many additional important observations on factorizing models whose
relation to the physical principles of QFT, unlike the bootstrap-formfactor
program, is not yet settled. The meaning of the c-parameter outside the chiral
setting and ideas on its renormalization group flow as well as the various
formulations of the thermodynamic Bethe Ansatz belong to a series of
interesting observations whose final relation to the principles of QFT still
needs clarification.

\section{Ongoing research, results from operator algebra methods}

QFT has been enriched by a the powerful new concept of modular localization
which promises to revolutionize the task of (nonperturbative) classification
and construction of models. It provides an additional strong link between
two-dimensional and higher dimensional QFT and admits a rich illustration for
chiral and factorizing theories. In the following we comment on two such
ongoing investigations.

One is motivated by the recent discovery of the adaptation of Einsteins
classical principle of local covariance to QFT in curved spacetime. The
central question raised by this work ($\rightarrow$ (78) Algebraic approach to
quantum field theory) is if all models of Minkowski spacetime QFTs permit a
local covariant extension to curved spacetime and if not which models do. In
the realm of chiral QFT this would amount to ask if all Moebius-invariant
models are also $Diff(S^{1})$-covariant.

The second one concerns the operator algebraic interpretation of temperature
duality which includes the Verlinde relation as a special case. This requires
the elaboration of a chiral analog of the Osterwalder-Schrader Euclideanization.

\subsection{Spacetime symmetries from the relative positions of monades}

Localized operator algebras $\mathcal{A}(\mathcal{O})$ for spacetime region
$\mathcal{O}$ with a nontrivial causal disjoint $\mathcal{O}^{\prime}$ are
under very general conditions (for wedge-localized algebras and interval
localized algebras of chiral QFT no additional conditions need to be imposed)
isomorphic to a unique algebra whose special role was highlighted in
mathematical work by Connes and Haagerup and whose physical raison d'etre is
the inexorable vacuum polarization associated with relativistic localization.
It is quite surprising that the full richness of QFT can be encoded into the
relative position of a finite number of copies of this ``monade''\footnote{We
borrow this terminology from the mathematician and philosopher Gottfried
Wilhelm Leibnitz; in addition to its intended philosophical content (reality
created by relations between monades) it has the advantage of being much
shorter than the mathematical terminology ``hyperfinite type III$_{1}$
Murray-von Neumann factor''. Instead of ``a finite number of copies of the
(abstract) monade'', we will simply say ``a finite number of monades''} within
a common Hilbert space \cite{testing}. Chiral conformal field theory offers
the simplest theoretical laboratory in which this issue can be analysed.

If the modular group $\sigma_{t}^{\mathcal{B}}$ in a \textit{joint standard}
situation for an inclusion $\left(  \mathcal{A\subset B},\Omega\right)  $ of
two monades (which share one $\Omega)$ acts on the smaller algebra for $t<0$
as a one-sided compression $\sigma_{t}^{\mathcal{B}}(\mathcal{A}%
)\subset\mathcal{A,}$ the two modular unitaries $\Delta_{\mathcal{A,B}}^{it}$
generate a unitary representation of a positive energy spacetime
translation-dilation (Anosov) group with the commutation relation (Borchers,
Wiesbrock 1992/93)
\begin{equation}
Dil(\lambda)U(a)Dil^{\ast}(\lambda)=U(\lambda a),\,Dil(e^{-2\pi t}%
)=\Delta_{\mathcal{B}}^{it}%
\end{equation}
The geometrical picture which goes with this abstract modular inclusion is
$\mathcal{B}=\mathcal{A}(I)\supset\mathcal{A}(\check{I})=\mathcal{A}$ with the
two intervals $\check{I}\subset I$ having one endpoint in common so that the
modular group of the bigger one $\simeq$ $Dil_{I}$ (Moebius transformation
leaving $\partial I$ fixed) leaves this endpoint invariant and compresses
$\check{I}$ into itself by transforming the other endpoint of $\partial
I^{\prime}$ into $I^{\prime}.$ One can show that this half-sided modular
inclusion situation ($\pm hsm,$ $t\lessgtr0$) actually cannot result from any
other von Neumann type than copies of the monade.

The simplest way to obtain the full Moebius group as a symmetry group of a
vacuum representation from pure operator-algebraic data is to require that the
modular inclusion itself is standard which means that in addition the vacuum
$\Omega$\ is also standard with respect to the relative commutant
$\mathcal{A}^{\prime}\cap\mathcal{B}$ (the third monade)$\mathcal{.}$ The
associated geometric picture is that of two half-circles whose intersection is
a quarter circle \cite{testing}.

\begin{theorem}
\bigskip The observable algebras of chiral QFT are classified by standard hsm
of two monades.
\end{theorem}

The net of interval-indexed local observable algebras is obtained by applying
the Moebius group to the original monade $\mathcal{A}\,$or $\mathcal{B}$ and
the problem of classifying chiral models is reduced to a well-defined problem
in the theory of operator algebras.

Encouraged by this successful encoding of the vacuum symmetry group of chiral
theories into the relative position of monades, it is natural to ask whether
this algebraic encoding can be extended to the vacuum-changing part of
$Diff(S),$ which is what the principle of local covariance would require.
Certainly all of the afore-mentioned models permit this extension since they
possess an stress-energy tensor whose Fourier decomposition leads to the
unitary implementation of $Diff(S).$ The known counterexamples \cite{testing}
of models which are Moebius invariant but lack the full $Diff(S^{1})$
covariance can be excluded on the basis of two well-motivated quantum physical
properties: \textit{strong additivity} and the \textit{split property
}\cite{testing}. So it is natural to ask whether these local quantum physical
requirements guaranty the extension $Moeb\rightarrow Diff(S^{1}).$from the
global vacuum preserving Moebius invariance to local Diff(S) covariance. This
is indeed possible if and only if in addition to the vacuum there exist other
state vectors $\Phi$ which with respect to certain (multi-) local subalgebras
$\mathcal{A}(I)$ lead to standard pairs ($\mathcal{A}(I),\Phi$) whose modular
group is \textit{partially geometric}. For a presentation of this concept and
its role in the extension problem see \cite{testing}.

\begin{theorem}
For strongly additive Moebius-invariant chiral models which fulfill the split
property, the $Diff_{2}(S)$ covariance is equivalent to the existence of a
partially geometric non-vacuum vectors $\Psi$ such that the modular group of
($\mathcal{A}(I)\vee\mathcal{A}(J),\Psi$) acts as $Dil_{2}$ on $I\cup J.$
\end{theorem}

The 3-parametric group $Diff_{2}(S)$ results from $Dil_{2}$ by changing the
position of the fixpoints through the application of Moebius transformations
and defining the group generated by these generalized dilations. The analogous
use of the $k^{th}$ instead of the $2^{nd}$ power leads to $Dil_{k}$
restricted to a $k$-fold localized algebra; with k-fold localized intervals
placed into a more general positions one generates $Diff_{k}(S)$. With this
construction we have reached our aim to encode the geometric extension problem
to $Diff(S)$ into a local quantum physical requirement. Whether these required
modular properties for securing the existence of the non-vacuum preserving
part can also be encoded into an algebraic significant positioning of monades
is presently not known.

The problem of characterizing Poincar\'{e} (or conformal) invariant higher
dimensional QFTs in terms of a finite number of monades has a positive answer;
in this case the local covariance principle has however only been only checked
for the Weyl algebra as well as in the perturbative approach to QFT on curved
spacetime \cite{testing}. It is hard to imagine how one can combine quantum
theory and gravity without understanding first the still mysterious links
between spacetime geometry, thermal properties and relative positioning of
monades in a joint Hilbert space.

\subsection{Euclidean rotational chiral theory and temperature duality}

Euclidean theory associated with certain real time QFTs is a subject whose
subtle and restrictive nature has largely been lost in many contemporary
publications as a result of the ``banalization'' of the Wick rotation (for
some pertinent critical remarks see \cite{testing}). The mere presence of
analyticity linking real with imaginary (Euclidean) time, without establishing
the subtle reflection positivity (which is necessary\ to derive the real time
spacelike commutativity as well as the Hilbert space structure), is not of
much physical use; what is needed is an operator algebraic understanding of
the so-called Wick rotation.

The issue of understanding Euclideanization in chiral theories became
particularly pressing after it was realized that Verlinde' observation on a
deep structural connection between fusion rules and modular transformation
properties of characters of irreducible representations of chiral observable
algebras is best taken care of by considering it as a part of a wider setting
involving angular parametrized thermal n-point correlation functions in the
superselection sector $\rho_{\alpha}$
\begin{align}
\left\langle A(\varphi_{1},..\varphi_{n})\right\rangle _{\rho_{\alpha}%
,2\pi\beta_{t}}  &  :=tr_{H_{\rho_{\alpha}}}e^{-2\pi\beta_{t}\left(
L_{0}^{\rho_{\alpha}}-\frac{c}{24}\right)  }\pi_{\rho_{\alpha}}(A(\varphi
_{1},..\varphi_{n}))\,\\
A(\varphi_{1},..\varphi_{n})  &  =\prod_{i=1}^{n}A_{i}(\varphi_{i})\nonumber
\end{align}
i.e. the Gibbs trace at inverse temperature $\beta=2\pi\beta_{t}$ on
observable fields in the representation $\pi_{\rho_{\alpha}}.$ Such thermal
states are (in contrast to the previously used ground states) independent on
the particular localization $loc\rho_{\alpha},$ they only depend on the
equivalence class i.e. on the sector $\left[  \rho_{\alpha}\right]
\equiv\alpha.$ These correlation functions\footnote{The conformal invariance
actually allows a generalization to complex Gibbs parameters $\tau$ with
$Im\tau=\beta$ which is however not neede in the context of the present
discussion.} fulfill the following thermal duality relation
\begin{equation}
\left\langle A(\varphi_{1},..\varphi_{n})\right\rangle _{\alpha,2\pi\beta_{t}%
}=\left(  \frac{i}{\beta_{t}}\right)  ^{a}\sum_{\gamma}S_{\alpha\gamma
}\left\langle A(\frac{i}{\beta_{t}}\varphi_{1},..\frac{i}{\beta_{t}}%
\varphi_{n})\right\rangle _{\gamma,\frac{2\pi}{\beta_{t}}} \label{duality}%
\end{equation}
where the right hand side formally is a sum over thermal expectation at the
inverse temperature $\frac{2\pi}{\beta_{t}}$ at the analytically continued
pure imaginary values scaled with the factor $\frac{1}{\beta_{t}}.$ The
multiplicative scaling factor in front which depends on the scaling dimensions
of the fields and is just the one which one would naively write if the
transformation $\varphi\rightarrow\frac{i}{\beta_{t}}\varphi$ were an
admissible conformal transformation law.

This relation can be checked explicitly (using Poisson resummation techniques)
in simple models as the abelian current models \cite{testing}. Since the Gibbs
states are not normalized, the Kac-Peterson-Verlinde character identities are
actually the ``zero-point function'' part (i.e. $A=1)$ of the above relation
(with the statistics character matrix $S$ already mentioned at the end of
section 4).A model-independent derivation of \ (\ref{duality}) can be given in
the operator algebraic setting of \textit{angular Euclideanization. }This
theory leads to a map which takes a dense analytic subalgebra of
$\mathcal{A}(S^{1})$ into one of a ``Euclidean'' theory which apart from
having a different Hilbert space inner product and consequently a different
star operation but for which the respective closures are analogous. In the
ensuing identity between the correlation functions of pointlike covariant
generators (\ref{duality}) the statistics character matrix S enters in an
intersting way and together with another diagonal phase matrix $T$ leads to a
situation in which the discrete modular group $SL(2,R)$ plays the role of a
new $SL(2,R)$ symmetry-like structure \cite{testing}.

\end{document}